# Two-Photon Dual-Comb LiDAR

Hollie Wright (1), Jinghua Sun (1,2), David McKendrick (3), Nick Weston (3) and Derryck T. Reid (1,*)

*(1) Scottish Universities Physics Alliance (SUPA), Institute of Photonics and Quantum Sciences, School of Engineering and Physical Sciences, Heriot-Watt University, Edinburgh EH14 4AS, UK.*
*(2) School of Electronic Engineering and Intelligentization, Dongguan University of Technology, Dongguan, Guangdong 523808, China.*
*(3) Renishaw Plc, Edinburgh, EH14 4AS, UK.*
*\*D.T.Reid@hw.ac.uk*

**Abstract:** The interferometric signals produced in conventional dual-comb laser ranging require femtosecond lasers with long-term $f_{CEO}$ stability, and are limited to an upper sampling rate by radio-frequency aliasing considerations. By using cross-polarized dual combs and two-photon detection we demonstrate carrier-phase-insensitive cross-correlations at sampling rates of up to 12× the conventional dual-comb aliasing limit, recording these in a digitizer-based acquisition system to implement ranging with sub-100-nm precision. We then extend this concept to show how the high data burden of conventional dual-comb acquisition can be eliminated by using a simple microcontroller as a ns-precision stopwatch to record the time intervals separating the two-photon cross-correlation pulses, providing real-time and continuous LiDAR-like distance metrology capable of sub-100 nm precision and dynamic acquisition for unlimited periods.

## 1. Introduction

Cross-correlation time-of-flight metrology employs two frequency combs, identical in all respects except for a small repetition-frequency difference of $\Delta f_{rep}$. One comb probes the distance to be measured using target and reference pulse sequences, while the other serves as a local oscillator (LO) that temporally gates the arrival of these pulses on a detector. The technique is a simplification of phase-coherent dual-comb metrology [1], and has been demonstrated with combs whose carrier-envelope offset frequencies are uncontrolled [2,3]; the resulting interferograms can be Hilbert-transformed to provide envelopes whose positions are localized by a Gaussian-fitting procedure. While the use of free-running combs substantially simplifies the technique, the unavoidable long-term drift in their carrier-envelope offset frequencies ($f_{CEO}$) makes the carrier frequency of the interferograms unstable, causing Hilbert-transform processing to fail when the frequency is too low. While stable interferograms can be ensured by using fully-locked frequency combs [4,5,6], this leads to a complex system requiring four locking loops (two for $f_{rep}$ and two for $f_{CEO}$). Previously, distance metrology that was insensitive to variations in $f_{CEO}$ was implemented by cross-correlating orthogonally polarized probe and LO pulses in two Type II BBO crystals [7]. A peak-fitting algorithm was applied to the digitized cross-correlation traces to extract the target-reference path distance. The related balanced optical cross-correlation technique was implemented using a periodically-poled potassium titanyl phosphate (PPKTP) crystal [8,9], providing a signal whose zero-crossings could be used to localize the coincidence of the LO and target / reference pulses. Both techniques achieved precisions comparable to the original dual-comb time-of-flight demonstration [2], although introduced greater complexity in the form of multiple detectors and nonlinear crystals.

An alternative to schemes using nonlinear crystals is two-photon cross-correlation, which is already well established as a method for optical pulse measurement [10,11] and offers multiple benefits over an interferometric dual-comb signal. Similar to cross-correlation in Type II crystals, the non-interferometric nature of the two-photon cross-correlation signal removes the need to phase-stabilize the participating lasers. Normally dual-comb metrology requires tight optical phase control of the two combs so that the interferogram carrier frequency is stable. When the carrier-envelope offset frequencies of the lasers are not actively stabilized, even tiny environmental changes can cause the interferogram carrier frequency to fluctuate rapidly across the full DC to $f_{rep}/2$ range, again making envelope extraction problematic. Two-photon cross-correlation is immune to this effect, allowing the use of lasers with free-running $f_{CEO}$ frequencies.

Unlike interferometric dual-comb methods, the lasers can also operate at different wavelengths without impacting the two-photon signal, and enabling the constraints on the laser design and performance to be relaxed. Since the two-photon absorption process is not phasematched, wavelength constraints are relaxed compared to schemes that use upconversion in nonlinear crystals [7-9]. By utilizing a wider bandgap detector than conventional dual-comb metrology, two-photon cross-correlation is also insensitive to light at the laser wavelength, except where this takes the form of short pulses focused onto the detector. Critically, these constraints provide rejection of stray light and amplified spontaneous emission from the metrology signal, both potentially important benefits where an optical amplifier is used in the receiver channel of a dual-comb system.

Perhaps the most significant advantage of utilizing a non-interferometric signal is the elimination of the conventional dual-comb aliasing condition, which places an upper limit on the repetition-frequency difference between the two lasers. The interference between two combs of repetition rates $\approx f_{rep}$, and sharing a common optical frequency bandwidth of $\Delta\nu$, creates radio frequencies which uniquely map onto optical frequencies only when the maximum repetition-rate difference $\Delta f_{rep} < f_{rep}^2/2\Delta\nu$. As this limiting value is approached, the interferogram contains frequencies approaching DC, which modulate its shape and make time-of-flight analysis from envelope extraction unreliable. By contrast, no equivalent limitation applies to two-photon cross-correlation. The value of $\Delta f_{rep}$ determines the metrology sampling rate, so the ability to exceed the conventional aliasing limit enables faster data acquisition, facilitating dynamic measurements and potentially improving the measurement precision in any given averaging time.

A further advantage for metrology is that the duration of the two-photon cross-correlation signal is significantly less than that of the conventional dual-comb interferogram. As described in [12], the second-order intensity cross-correlation is shorter than the first-order field cross-correlation by a factor of 0.707 for Gaussian pulses, and by a factor of 0.625 for $\text{sech}^2(t)$ pulses of the kind commonly produced by soliton lasers. This considerable difference makes the two-photon cross-correlation signal particularly well suited as a signal for time-of-flight metrology, potentially enabling more precise localization of its center, and providing a more sharply rising edge for LiDAR-type timing measurements.

Here, we introduce an experimental embodiment of two-photon cross-correlation dual-comb ranging. We demonstrate acquisition well beyond the conventional dual-comb aliasing limit by directly comparing the two-photon signals with conventional one-photon dual-comb interferograms. Using the standard digitizer-based acquisition approach we obtain sub-100-nm precision in ≈2 seconds at an update rate of 2.4× the conventional dual-comb aliasing limit. Finally, we show how the digital nature of the two-photon signal can be used to configure a LiDAR-like time-of-flight system that provides dynamic absolute distance measurement with few-µm instantaneous precision.

## 2. Distance metrology using two-photon cross-correlation

*2.1 Laser and metrology system*

The combs, which each followed the layout illustrated in Fig. 1(a), were SESAM-modelocked Er:fiber lasers producing pulses with durations of 418 fs and 716 fs, repetition rates of $f_{rep} \approx$ 78.87 MHz and center wavelengths close to 1555 nm. The lasers were based on previously reported designs [13,14] and incorporated either fiber butt-coupling or free-space focusing onto a semiconductor saturable absorber mirror (SESAM), as described in [15,16]. Both cavities comprised only polarization-maintaining (PM) fibers. The lasers were each terminated by a piezo-actuated 98% reflectance output coupler (OC), which in the LO comb was mounted on a translation stage to allow coarse tuning of the repetition frequency. The output-coupled power was approximately 900 µW from each laser oscillator. A short free-space section was used to couple the light from each comb into a two-stage, non-PM Er:fiber amplifier, which boosted the probe and LO laser powers to around 40 mW in each channel.

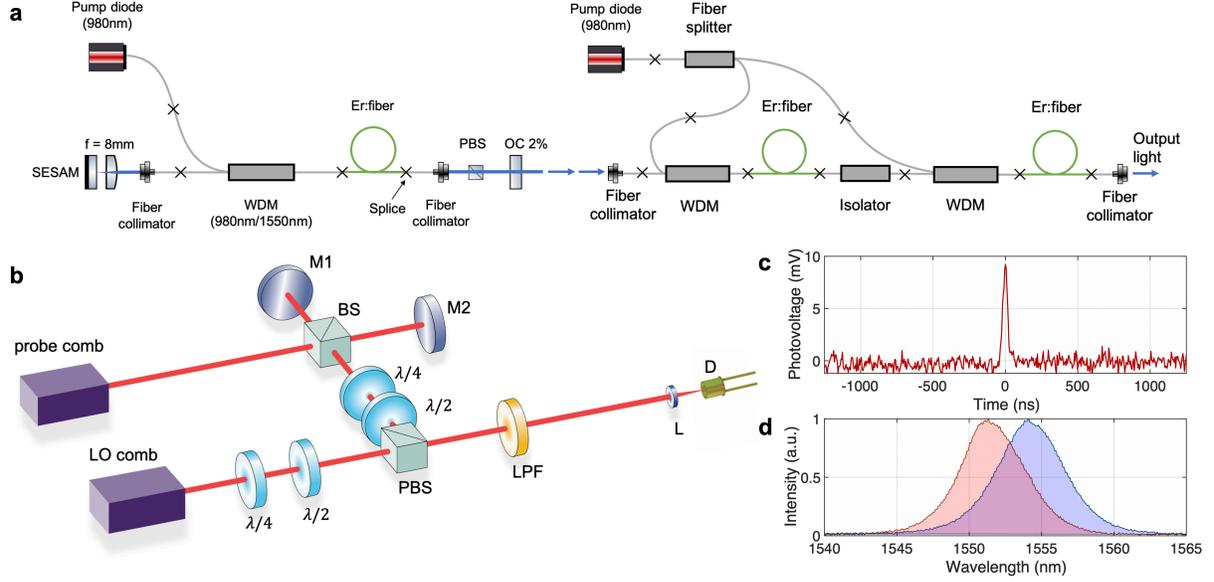

Fig. 1. (a) Design of the Er:fiber laser combs, in which a low-power master oscillator is free-space coupled into a two-stage amplifier with total gain of 16.5 dB. One comb was constructed to exactly this design, while the other was identical except for the use of a fiber-butt-coupled SESAM instead of free-space coupling. (b) Dual-comb two-photon-absorption (TPA) detection scheme. BS: non-polarizing beamsplitter; PBS: polarizing beamsplitter; D: silicon avalanche photodetector; LPF: long-pass filter ($\lambda >1300$ nm). (c) Representative two-photon cross-correlation signal. (d) Optical spectra of the probe (blue) and LO combs (red).

As shown in Fig. 1(b), pulses from the probe comb were split into two replicas by a beamsplitter, which were reflected from reference and target mirrors before being recombined. The recombined pulses were prepared in an *s*-polarization by using quarter-wave and half-wave plates. The pulses from the LO comb were made *p*-polarized in a similar way before being combined with the probe pulses at a polarizing beamsplitter. In this way, the probe and LO pulses were combined with orthogonal polarizations, ensuring that no interference could occur. The combined beams passed through a long-pass filter to block unabsorbed 980-nm pump light before being focused onto a silicon avalanche photodiode (-3-dB bandwidth of 100 MHz) to produce a two-photon intensity cross-correlation signal. The signal was low-pass filtered to reject frequencies above $f_{rep}/2$, before being acquired on a digitizer. Figure 1(c) shows an example of the signal obtained in this way. The system also included the option to simultaneously record a conventional dual-comb interferogram on an InGaAs PIN photodiode, which was supplied by light from a beamsplitter inserted immediately before the focusing lens of the silicon detector and used a 45° polarizer to resolve a common polarization component of the probe and LO pulses.

In order to provide the first harmonic of the comb repetition rates, the outputs of both lasers were sampled independently using further InGaAs detectors. The repetition rate of the LO comb was measured using a frequency counter for use in subsequent data analysis. The repetition-rate signals were mixed and low-pass filtered to obtain $\Delta f_{rep}$, which provided the trigger signal to the digitizer. This trigger signal was not strictly necessary but facilitated data analysis by ensuring the reference signal always appeared at the start of the digitized signal. The lasers were independently locked to stabilize $\Delta f_{rep}$ to a value of up to several kHz.

*2.2 Comparison of dual-comb signals using conventional and two-photon detection*

As discussed in Section 1, two-photon cross-correlation can facilitate metrology at update rates higher than the conventional dual-comb aliasing limit. The 15-nm common bandwidth of our combs (see Fig. 1(d)) leads to a requirement that, to avoid aliasing artefacts, the maximum repetition rate difference should be $\Delta f_{rep} < 1.67$ kHz. By contrast, the two-photon cross-correlation measurement encounters no such restriction. In Fig. 2 we present, on a common time axis, conventional and two-photon signals acquired simultaneously for values of $\Delta f_{rep}$ from 500 Hz–20 kHz. The corruption of the conventional dual-comb interferogram is clear to see in Figs. 2(e)–2(h), with a clearly defined envelope shape being lost entirely at the highest value of $\Delta f_{rep}$. Conversely, the corresponding two-photon cross-correlation signal remains clearly distinguishable for all values of $\Delta f_{rep}$, and Figs. 2(i)–2(p) illustrate how the width of the signal scales inversely with $\Delta f_{rep}$. As can be seen from Fig. 2(p), even sampling at $\Delta f_{rep} = 20$ kHz, corresponding to 12 the conventional aliasing limit, yields a clear two-photon cross-correlation signal on an oscilloscope. At first this result seems counterintuitive; the LO pulses gate the probe pulses at time intervals of $\Delta f_{rep}/f_{rep}^2$, which for pulse durations of $\Delta\tau$, appears to place a limit of $\Delta f_{rep} < f_{rep}^2 \Delta\tau$ if coincidence between the pulses is to be guaranteed. For transform-limited pulses, where $\Delta\tau\Delta\nu \approx 0.5$, this temporal limit is, to a first approximation, the same as the conventional aliasing limit, but two factors extend the limit in practice. Firstly, if the pulses reaching the detector are not transform-limited then their temporal overlap is increased, and the criterion is relaxed. Secondly, and similar to the operation of a sampling oscilloscope, when $f_{rep} \neq m\Delta f_{rep}$ (where $m$ is an integer) the cross-correlation trace will become better resolved with time, as eventually all possible relative delays between LO and probe pulses are sampled. These differences mean that no fundamental limit on $\Delta f_{rep}$ exists, although trade-offs are made between sampling-rate, precision and averaging time as the conventional aliasing limit is exceeded.

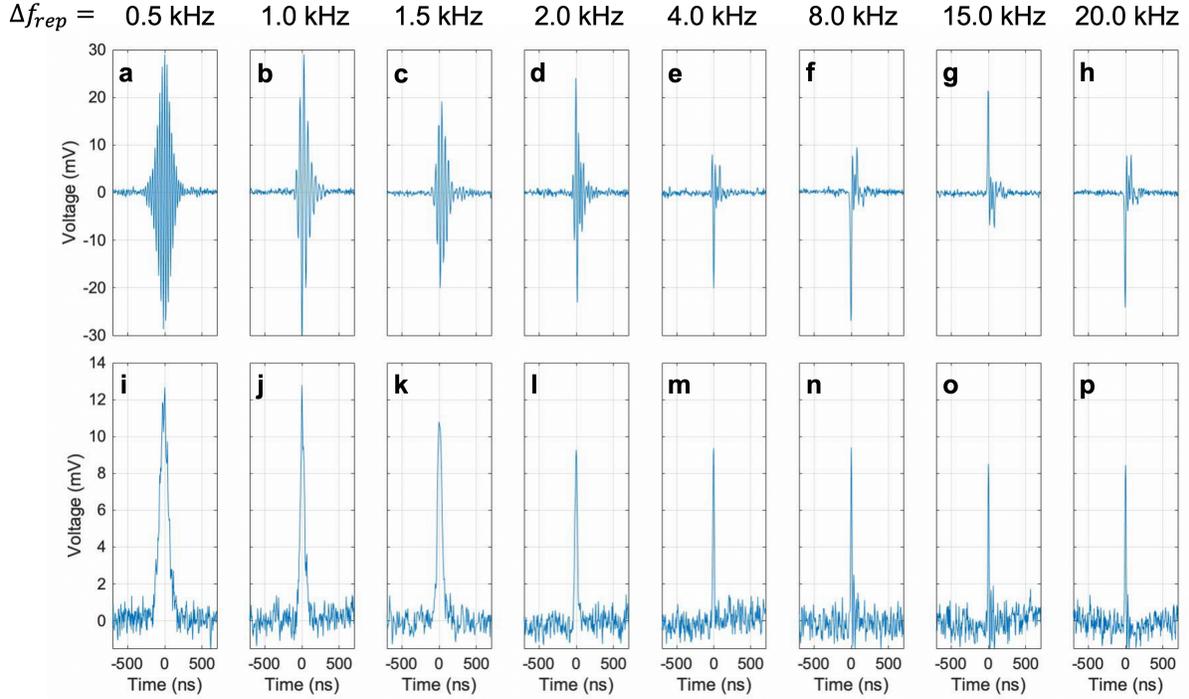

Fig. 2. (a)–(h): Conventional dual-comb interferograms collected for $\Delta f_{rep}$ values of 500 Hz, 1.0 kHz, 1.5 kHz, 2.0 kHz, 4.0 kHz, 8.0 kHz, 15.0 kHz and 20.0 kHz respectively. (i)–(p): Two-photon cross-correlation signals acquired simultaneously with the interferograms in (a)–(h). All data were acquired at a sampling rate of 200 MSa/s.

*2.3 Demonstration of time-of-flight ranging using two-photon detection*

We evaluated the performance of dual-comb ranging using the target-reference mirror scheme shown in Fig. 1 and with a roundtrip target-reference path difference of 24 cm. The repetition-frequency difference was set to $\Delta f_{rep} = 4$ kHz, corresponding to 2.4× the conventional aliasing limit. The two-photon cross-correlation signals were acquired on a GaGe Razormax digitizer, and an example of a short section of one dataset is shown in Fig. 3(a), with the more intense signals being from the target ($T$) mirror and the weaker signals from the reference ($R$) mirror. With reference to Fig. 3(a), the time-of-flight path difference for measurement $i$ is calculated as $v_g t_i^{TR}/(t_i^{TR} + t_i^{RT}) f_{rep(LO)}$, where the signal period is $t_i^{TR} + t_i^{RT} = 1/\Delta f_{rep}$. The Allan deviation of the distance measurement is shown in Fig. 3(b) for averaging times of up to two seconds. For the minimum acquisition time of 250 μs (set by the 4-kHz difference in the repetition rates) the measurement precision is 17.8 μm. As the fitted $\sqrt{\tau}$ line shows, the precision scales as the square root of the number of samples, reaching sub-μm precision for 0.1 s of averaging. After 1 s of averaging, the precision reaches 150 nm ($\lambda/10$), falling to 93 nm after 2 s of averaging, which is below the $10^{-7}$ uncertainty limit imposed by atmospheric fluctuations. This performance is comparable with previously published results from conventional dual-comb distance metrology, which achieve few-μm precisions in around 1 ms averaging times [2,3,7,8,9].

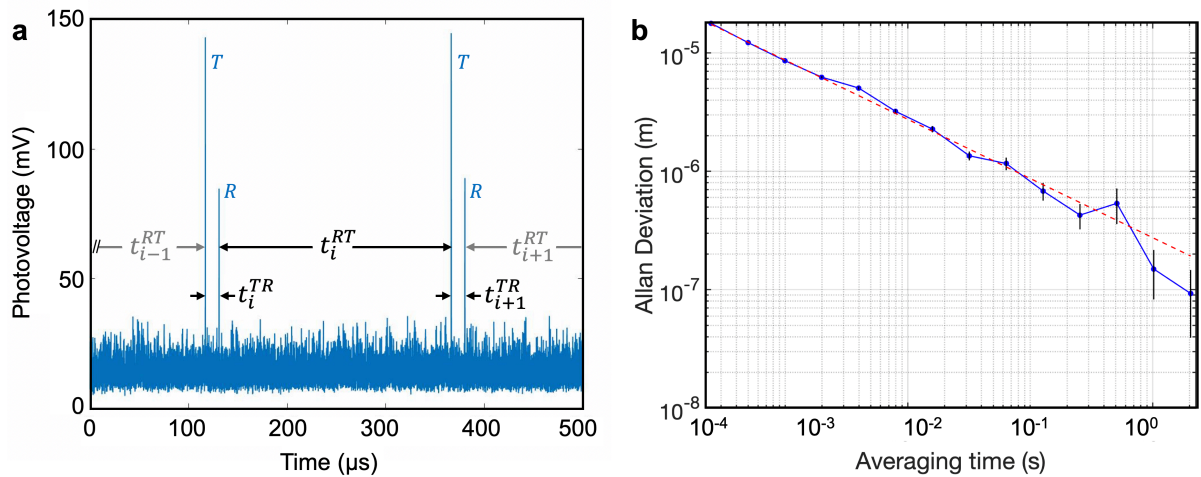

Fig. 3 (a) Two-photon dual-comb cross-correlations between the local oscillator and the target ($T$) and reference ($R$) reflections, recorded over a time of nearly two measurement periods ($2/\Delta f_{rep}$). The notation defines $t_i^{TR}$ as the time delay of the $i^{th}$ LO-reference cross-correlation after the previous LO-target cross-correlation, and $t_i^{RT}$ as the delay of the $i^{th}$ LO-target cross-correlation after the previous LO-reference cross-correlation. (b) Analysis of the digitizer dataset containing the measurements presented in (a), showing an Allan deviation with an instantaneous precision of 17.8 μm, falling to 150 nm for 1s of averaging. The dashed red line has a slope of -0.5.

## 3. Extension to two-photon dual-comb LiDAR

In conventional LiDAR, the time-of-flight of a tens-of-nanosecond pulse reflected from a target is measured directly using electronics with sub-100-ps timing resolution, equating to few-cm precision. The digital character of the two-photon cross-correlation signals allows them to be used in a similar way, but with the important distinction that the true time-of-flight is magnified by a factor of $f_{rep}/\Delta f_{rep}$ – a value of between 20,000 and 400,000 in our system. Even using timing electronics with only modest resolution, this magnification factor allows few-μm precision to be achieved directly from dynamic timing signals, circumventing the extreme data burden which can be encountered in dual-comb techniques which digitize the full interferometric signal.

In a proof-of-concept experiment, we used a simple microcontroller as a ns-precision stopwatch to record the time intervals separating the cross-correlations between the LO pulses and successive target and reference pulses. The microcontroller was operated with a clock frequency of 600 MHz (cycle period 1.67 ns), and was configured to continuously count clock cycles in an onboard register. A rising edge presented on one of the pins triggered an interrupt which reset the cycle count and wrote the previous total count to a serial interface for use by other software.

We conditioned the two-photon cross-correlation pulses from their raw 10-mV peak-peak format by first amplifying them to around 500 mV peak-peak using an inverting RF amplifier (Figs. 4(a) and 4(b)), following which further amplification and a two-stage Schmitt trigger were used to provide clean 3.3 V pulses suitable for triggering the microcontroller (Fig. 4(c)).

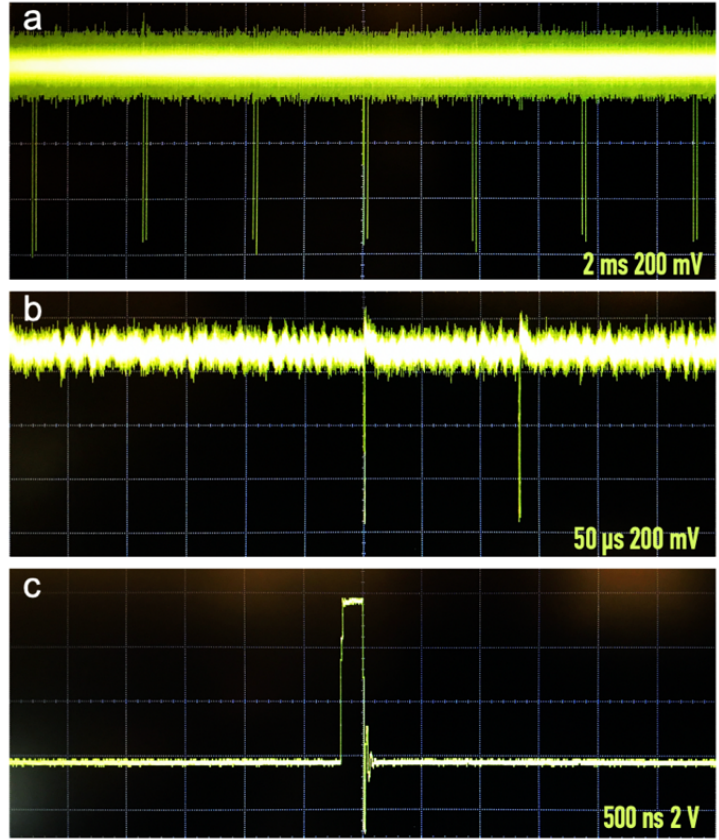

Fig. 4(a) Oscilloscope images showing pairs of two-photon cross-correlations between the LO pulses and successive target / reference pulses after first-stage amplification, and (b) a single pair of target / reference cross-correlations. (c) Example of a single cross-correlation after the Schmitt trigger circuit. After attenuation this was used as the input to the microprocessor.

The resulting signal comprised pairs of short electrical pulses, with one pulse in each pair encoding the target reflection ($T$) and the other pulse encoding the reference reflection ($R$). With reference to the notation in Fig. 3, the time intervals between the pairs of pulses can be denoted as, $t_i^{TR}$ and $t_i^{RT}$. The counter, which increments once every microcontroller clock period ($\tau_{CPU}$), is reset by the arrival of each pulse. When a reset occurs, the microcontroller sends the previous counter value to the serial bus, and in this way a series of alternating numbers is recorded, $n_i^{TR} = t_i^{TR}/\tau_{CPU}$ and $n_i^{RT} = t_i^{RT}/\tau_{CPU}$, which can be used to obtain the instantaneous absolute distance using the probe-laser pulse repetition frequency, $f_{rep}$, according to:

$$d_i = \frac{v_g}{2f_{rep}} \cdot \frac{n_i^{TR}}{n_i^{TR} + n_i^{RT}} \tag{1}$$

The instantaneous repetition-rate difference between the combs is similarly available from the timing signals using:

$$\Delta f_{rep,i} = \frac{1}{\tau_{CPU}(n_i^{TR} + n_i^{RT})} \tag{2}$$

The time $t_j$ at which distance $d_j$ is measured is computed from the counter values:

$$t_j = \sum_1^j \tau_{CPU}(n_i^{TR} + n_i^{RT}) \tag{3}$$

Extracting the lab time in this way makes the measurement insensitive to changes in the repetition-frequency difference between the lasers, allowing for accurate measurements of dynamic changes in an object's position.

An illustration of the approach is shown in Fig. 5. Data in Figs. 5(a)–5(c) were obtained with static target and reference mirrors when the combs were locked with $\Delta f_{rep} = 2430$ Hz. The dataset length of around 25 seconds occupied only a few 100 kB, corresponding to many orders of magnitude less than the equivalent for digitizer-based measurements. Figure 5(a) shows the CPU cycle counts recorded, and the inset (1 ms/div scale) illustrates the recorded data points. A distance of 47.832 mm was inferred using Eqn. (1) and its real-time measurement is shown in Fig. 5(b); the data have a standard deviation of $\pm 12$ μm, broadly matching the distance uncertainty determined by one clock cycle of $v_g \tau_{CPU} \Delta f_{rep}/f_{rep}$. By using Eqn. (2) the instantaneous value of $\Delta f_{rep}$ can be determined. Since for this measurement the comb repetition rates were both locked, the obtained value of $\Delta f_{rep}$ is almost constant at 2430 Hz, however small exponentially-decaying oscillations can be seen (Fig. 5(c), arrows) that reveal the comb locking electronics responding to environmental perturbations. Finally, in Fig. 5(d) we show a series of distance measurements, where a mirror mounted on a loudspeaker was driven sinusoidally at 2 Hz for increasing amplitudes. Data are shown for loudspeaker amplitudes of up to 375 μm and are each offset vertically by 1 mm for clarity.

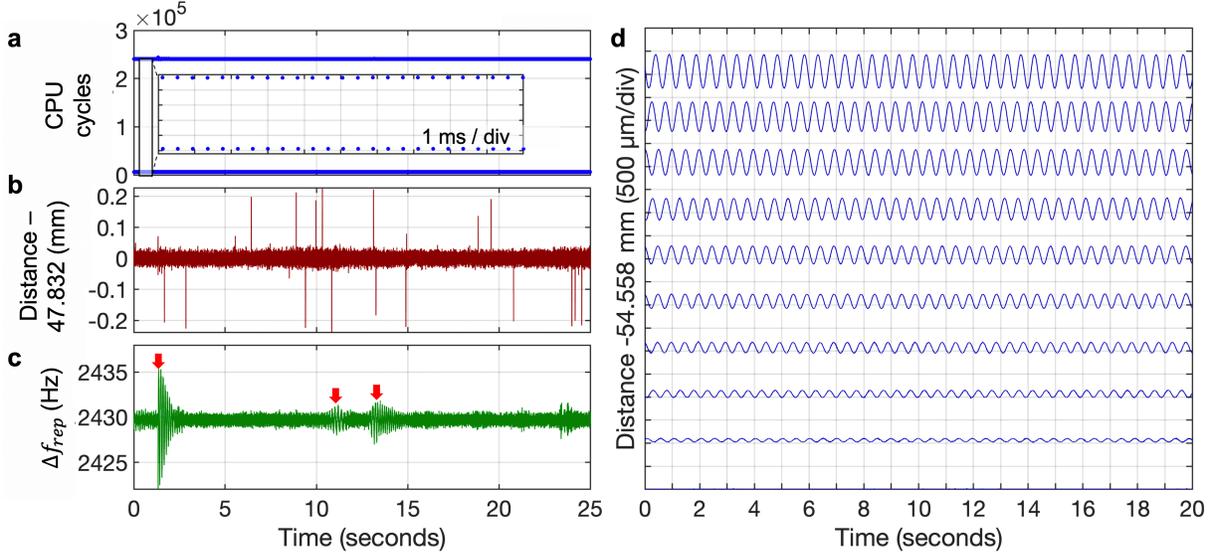

Fig. 5. (a) Counted cycles over a 25-second interval. The inset shows, respectively, the counts $n_i^{RT}$ and $n_i^{TR}$ over a 10-ms interval. (b) The absolute distance, extracted using Eq. (1). (c) The instantaneous value of the repetition-rate difference, $\Delta f_{rep,i}$, computed from Eq. (2) and shown on a timebase derived using Eq. (3). The arrows indicate small changes in $\Delta f_{rep}$ caused by environmental perturbations to one comb, and the consequent reaction of the locking electronics, which leads to a damped oscillation that restores the comb to the locked condition. (d) Dual-comb LiDAR measurement of a 2-Hz sinusoidally scanning mirror. Multiple datasets were recorded for increasing mirror amplitudes and are shown with a vertical offset of 1 mm. The measurement for zero amplitude lies under the time axis.

We assessed the precision of the absolute distance measurement obtained using this timer-based approach using an Allan deviation analysis, shown in Fig. 6, of the distance recorded for a static target mirror. The data, again recorded with $\Delta f_{rep} = 2430$ Hz, show similar precisions to those obtained using full digitizer-based sampling, achieving sub-μm precision in 50 ms, and sub-100-nm precision in around 3 s.

By using dedicated time-of-flight electronics we expect to be able to further improve the precision. Commercially available time-of-flight modules can achieve a timing resolution of 55 ps [17], which for $\Delta f_{rep} = 2430$ Hz and $f_{rep} \approx 78$ MHz implies a timing-limited distance precision of 250 nm.

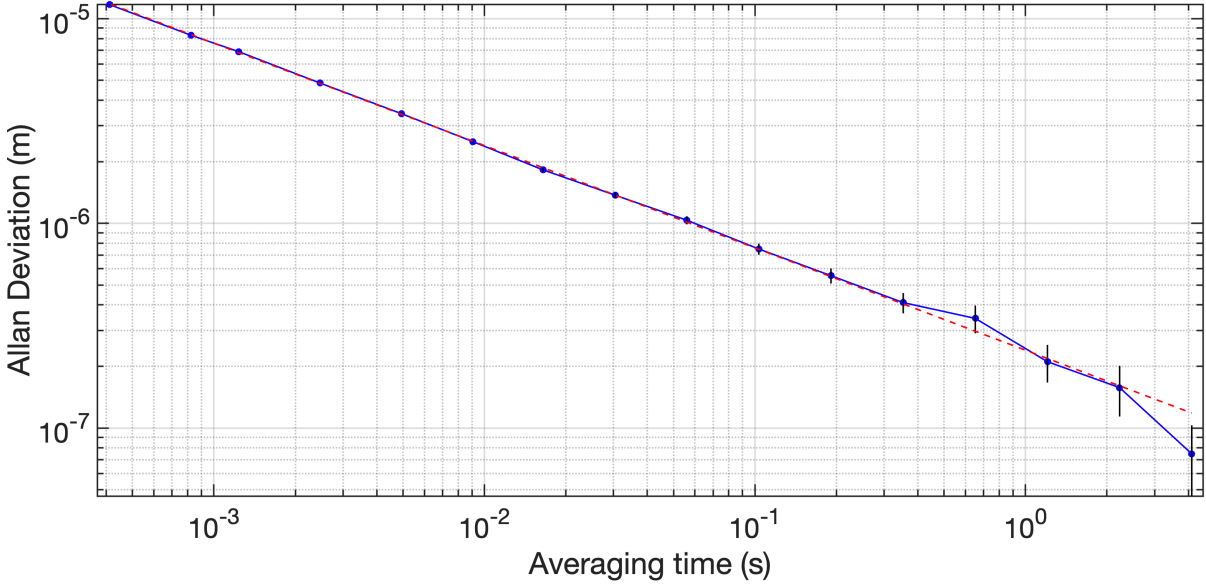

Fig. 6. Allan deviation of the absolute distance measurements obtained from approximately 20 seconds of data. The distance is calculated using Eq. (1) from a continuous stream of cycle counts provided by the microcontroller. The dashed red line has a slope of -0.5.

## 4. Conclusions

Two-photon dual-comb cross-correlation time-of-flight metrology shows considerable potential for dynamic, absolute distance measurements. The non-interferometric nature of the measurement is unconstrained by the conventional dual-comb aliasing limit, allowing operation at considerably higher acquisition rates than interferometric dual-comb metrology. Polarisation, wavelength and $f_{CEO}$ stability constraints on the lasers are considerably relaxed or removed entirely, while two-photon detection suppresses any contributions to the signal from ambient light or amplified spontaneous emission. The short, transient character of the two-photon signal is well suited to LiDAR-type timer-based detection, removing the vast data overhead normally associated with dual-comb techniques and allowing for dynamic and continuous absolute distance monitoring at few-µm, and potentially sub-µm resolution.

**Funding**

UK Engineering and Physical Sciences Research Council (EPSRC) grants EP/L01596X/1 and EP/N002547/1; Renishaw PLC.

**Acknowledgement**

We are grateful to Prof. Ron Hui and his group at the University of Kansas for early influential discussions on the Er:fiber laser design.

**Disclosures**

The authors declare no conflicts of interest.

**Data Availability Statement**

Data underlying the results presented in this paper are not publicly available at this time but may be obtained from the authors upon reasonable request.